\documentclass[pdflatex,sn-mathphys-num]{sn-jnl}%

\usepackage{graphicx}%
\usepackage{multirow}%
\usepackage{amsmath,amssymb,amsfonts}%
\usepackage{amsthm}%
\usepackage{mathrsfs}%
\usepackage[title]{appendix}%
\usepackage{xcolor}%
\usepackage{textcomp}%
\usepackage{manyfoot}%
\DeclareNewFootnote{A}[gobble]
\makeatletter
\newcommand{\equalcontribmark}{\g@addto@macro\artauthors{$^{\dagger}$}}
\makeatother

\makeatletter
\patchcmd{\@maketitle}
  {\removelastskip\vskip24pt}
  {\removelastskip\vskip10pt}
  {}{\PackageWarning{frontmatter}{Could not tighten first title-page gap}}
\patchcmd{\@maketitle}
  {\removelastskip\vskip24pt}
  {\removelastskip\vskip10pt}
  {}{\PackageWarning{frontmatter}{Could not tighten second title-page gap}}

\usepackage{booktabs}%
\usepackage{algorithm}%
\usepackage{algorithmicx}%
\usepackage{algpseudocode}%
\usepackage{listings}%

\usepackage{subcaption}
\usepackage{bbm}
\usepackage{bm}

\theoremstyle{thmstyleone}%

\theoremstyle{thmstyletwo}%

\theoremstyle{thmstylethree}%

\newcommand{\revisefoot}[1]{\footnote{\color{blue} #1}}
\renewcommand{\revisefoot}[1]{} 

\begin{document}

\title[Marginal Data Augmentation for Efficient Bayesian Modeling of Counts and Rates]{Marginal Data Augmentation for Efficient Bayesian Modeling of Counts and Rates {with a Demographic Application}}

\author*[1]{\fnm{Gregor} \sur{Zens}}\equalcontribmark
\author[2]{\fnm{Sylvia} \sur{Frühwirth-Schnatter}}\equalcontribmark

\affil[1]{\orgname{International Institute for Applied Systems Analysis}, \orgaddress{\street{Schlossplatz 1}, \city{Laxenburg}, \postcode{2361}, \state{Lower Austria}, \country{Austria}}}

\affil[2]{\orgdiv{Institute for Statistics and Mathematics}, \orgname{WU Vienna University of Economics and Business}, \orgaddress{\street{Welthandelsplatz 1}, \city{Vienna}, \postcode{1020},
\country{Austria}}}

\artnote{%
$^{\dagger}$Both authors contributed equally to this work. $^{*}$\textbf{Corresponding author:} Gregor Zens, \href{mailto:zens@iiasa.ac.at}{\nolinkurl{zens@iiasa.ac.at}}; Sylvia Frühwirth-Schnatter, \href{mailto:sfruehwi@wu.ac.at}{\nolinkurl{sfruehwi@wu.ac.at}}.  \textbf{This work has been published as:} Zens, G., Frühwirth-Schnatter, S. (2026). \emph{Marginal Data Augmentation for Efficient Bayesian Modeling of Counts and Rates with a Demographic Application}. In: Nagler, T., Kurowicka, D., Cooke, R., Joe, H. (eds) \emph{Statistical Dependence Modeling}. Springer, Cham. \href{https://doi.org/10.1007/978-3-032-14252-8_15}{\nolinkurl{https://doi.org/10.1007/978-3-032-14252-8_15}}.}

\abstract{Count data models are ubiquitous in many fields, yet Bayesian data augmentation algorithms for such models frequently encounter challenges with Markov chain Monte Carlo efficiency. Posterior simulation is especially demanding when modeling data with a high proportion of zero outcomes. In this paper, we address this issue by introducing a marginal data augmentation approach for semi-parametric Bayesian count data regression models, based on a working parameter that rescales latent outcomes corresponding to zero-count observations. This strategy alleviates the strong posterior dependencies that typically reduce the efficiency of standard data augmentation schemes and leads to substantial gains in sampling efficiency. Synthetic data examples and simulation studies are used to demonstrate the improvements in mixing compared to conventional sampling methods. A demographic application using latent factor analysis to model subnational mortality counts in Austria further underscores the broader applicability of the proposed methodology.}


\maketitle
\newpage
\section{Introduction}

\begin{figure}
    \centering
    \begin{subfigure}[b]{0.48\textwidth}
        \centering
        \includegraphics[width=\textwidth]{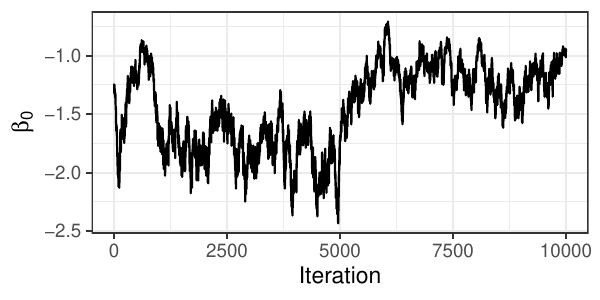}
        \caption{$\beta_0=-0.5$}
        \label{fig:y equals x}
    \end{subfigure}
    \hfill
    \begin{subfigure}[b]{0.48\textwidth}
        \centering
        \includegraphics[width=\textwidth]{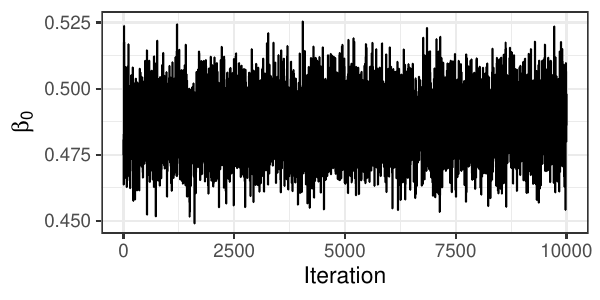}
        \caption{$\beta_0=0.5$}
        \label{fig:three sin x}
    \end{subfigure}
    \caption{Traceplots of $\beta_0$ obtained using a DA-MCMC algorithm for a STAR model, where $y_i =\lfloor\exp(z_i)\rfloor$, $z_i = \beta_0 + \epsilon_i$, and $\epsilon_i \sim \mathcal{N}(0, \sigma^2)$, see Sec.~\ref{sec:model} for details. In both scenarios, the true value of $\sigma^2$ is 0.05. The left panel illustrates the case with $\beta_0 = -0.5$, leading to $98\%$ zero counts and an inefficiency factor of $\text{IE}=861$. The right panel depicts a scenario with $\beta_0 = 0.5$, resulting in only $2\%$ zero counts and an inefficiency factor of $\text{IE}=3$.}
    \label{fig:trace_intro}
\end{figure}

Statistical models for counts and rates are fundamental tools in numerous research fields, including epidemiology, ecology, and demography. When working with count data, researchers face various challenges including overdispersion, zero-inflation, and complex dependency structures.\footnote{Claudia Czado has contributed significantly to this field, see \cite{gsc-cza:mod},  \cite{czado2009predictive}  and \cite{czado2005bayesian}, among other papers.} The Bayesian paradigm offers several advantages in addressing these challenges. For instance, the integration of spatio-temporal smoothing structures, zero-inflation mechanisms, or accounting for model uncertainty is straightforward based on suitable (hierarchical) prior specifications. From an inferential perspective, the Bayesian approach allows for fully probabilistic inference in complex count data models while simultaneously overcoming issues such as non-existence of maximum likelihood estimates (\cite{silva2010existence}), as proper prior densities can be used to regularize potentially degenerate likelihood functions.

Within the Bayesian paradigm, various computational techniques are available for posterior simulation, with Markov chain Monte Carlo (MCMC) algorithms traditionally playing a prominent role (\cite{gamerman1997sampling};  \cite{chib2001markov}; {\cite{pol-etal:bay_inf}}).
However, generic MCMC techniques such as Metropolis-Hastings samplers for generalized linear models can suffer from poor mixing and require substantial tuning efforts when targeting high-dimensional posterior densities.
A convenient alternative are data augmentation MCMC (DA-MCMC) algorithms (\cite{tanner1987calculation}) that introduce a set of latent outcomes in addition to the observed data in a way that facilitates posterior simulation. For complex and highly parameterized models, this is often the only feasible method for 
{achieving} (exact) posterior {inference}. 
DA-MCMC algorithms for count data have been proposed for models based on Poisson likelihoods (\cite{fruhwirth2006auxiliary}; \cite{fruhwirth2009improved}), negative binomial likelihoods (\cite{pillow2012fully}), Poisson log-normal and binomial logistic-normal models (\cite{steel2024model}) as well as for semi-parametric count data regression models (\cite{canale2013nonparametric}; \cite{kowal2020simultaneous}).

In practice, a key issue with DA-MCMC algorithms for count data is that they tend to mix poorly in settings with many zero outcomes. To illustrate this, Figure~\ref{fig:trace_intro} presents trace plots of the intercept parameter in a semi-parametric count data regression model applied to data sets with either large counts (right panel) or small counts, with over 98\% zero outcomes (left panel). MCMC efficiency deteriorates substantially in the presence of many zero counts. Importantly, this is not an issue related to a single specific DA-MCMC algorithm or a particular likelihood choice. Instead, this phenomenon affects many classical data augmentation schemes for semi-parametric models, as well as for models based on Poisson, binomial, or negative binomial likelihoods. Reasons for these inefficiencies will be discussed in more detail below.

To address these MCMC inefficiencies, this article proposes a \textit{marginal data augmentation} strategy (\cite{meng1999seeking}) for efficient estimation of a semi-parametric Bayesian count data regression model. A set of simulation studies is used to demonstrate that this novel MCMC scheme leads to substantial improvements in sampling efficiency, while requiring only minimal additional computational efforts. We further illustrate the practical significance of our methodology through a real-world demographic application. Building upon the pioneering work of Czado and colleagues (\cite{czado2005bayesian}), we develop a semi-parametric latent factor model for count data and apply it to Austrian subnational mortality data, highlighting the broader applicability of our approach.

The remainder of this article is structured as follows. Section \ref{sec:model} discusses a semi-parametric count data framework, as well as an MCMC algorithm for estimation. Section \ref{sec:px-da} introduces the proposed marginal data augmentation scheme. Section \ref{sec:simulations} presents results from simulation experiments based on synthetic data. Section \ref{sec:application} discusses a latent factor model application to subnational mortality data from Austria. Section \ref{sec:conclusion} provides a brief discussion of future research pathways and concluding remarks.

\section{Semi-Parametric Bayesian Modeling of Counts and Rates}
\label{sec:model}

\subsection{{Model definition}}

We consider a simple semi-parametric Bayesian count data regression model for non-negative integer counts $y_i \in \mathbb{N}_0$ for $i = 1, \dots, n$. The model is based on a latent Gaussian variable $z_i$
\begin{equation}
\label{eq:latent}
    z_i = \bm{x}_i^\top \bm{\beta} + P_i + \varepsilon_i \quad \text{with} \quad \varepsilon_i \sim \mathcal{N}(0, \sigma^2).
\end{equation}
which is assumed to be a linear function of an $R \times 1$ vector of observed covariates $\bm{x}_i$,  with an $R \times 1$ vector of coefficients $\bm{\beta}$ as well as a Gaussian error term $\varepsilon_i$ with mean zero and variance $\sigma^2$. In addition, we consider settings where an optional fixed and observed `offset' term $P_i$ may be included in the model. Such offset terms are, for instance, regularly used in epidemiological and demographic research, where they can be used to capture a varying \textit{exposure} or a varying \textit{population-at-risk} across observations $i$.

The latent outcome $z_i$ is related to the observed outcome $y_i$ via the choice equation
\begin{equation}
\label{eq:choice}
y_i \;=\;
\begin{cases}
0, & \text{if } z_i \in 
{(-\infty, 0)},
\\[6pt]
k, & \text{if } z_i \in
{\bigl[\log(k), \log(k+1)\bigr)}
,\ k \ge 1.
\end{cases}
\end{equation}

The model defined by (\ref{eq:latent}) 
and  (\ref{eq:choice}) is a special case of a class of models discussed in more detail in \cite{kowal2020simultaneous} and extends latent utility models typically used for binary and categorical data to integer-valued data. The boundaries in (\ref{eq:choice}) imply a logarithmic link function. We focus on this setting as the logarithmic link is the most commonly encountered link function in count data models, naturally respects non-negativity of the latent utilities after transforming, and allows for straightforward interpretation of regression parameter estimates. A similar framework where the boundaries of $z_i$ are defined via an identitiy link is considered in \cite{canale2013nonparametric}. The case of an unknown link function that is estimated from the data is discussed in \cite{kowal2020simultaneous}.

\subsection{{Bayesian Inference via  MCMC Sampling}}

{For Bayesian inference}, we assume a
semi-conjugate prior setup where $\bm{\beta} \sim \mathcal{N}(0, \bm{B}_0)$ and $\sigma^2 \sim \mathcal{IG}(c_0, C_0)$. {These priors are conjugate for the conditional inference problem where either  $\bm{\beta}$ or $\sigma^2 $ is known.}
In this case, all relevant full conditional densities are available in closed form, allowing a simple Gibbs sampler to be set up for posterior simulation. In addition, hierarchical priors on $\bm{B}_0$ allow to easily incorporate spatio-temporal dependence structures or sparsity in $\bm{\beta}$ via shrinkage priors, if desired.

 Conditional on {the latent variables $\bm{z}= (z_1,\ldots, z_n)$}, the
 model {reduces to}
 a Gaussian regression model, resulting in closed-form conditional posteriors of the remaining parameters. This makes data augmentation (\cite{tanner1987calculation}) a natural choice for posterior simulation. Instead of attempting to directly sample from {the posterior} $ p(\bm{\beta}, \sigma^2 \mid \bm{y})$ {given the data $\bm{y}=(y_1, \ldots,y_n)$}, which is typically difficult,  data augmentation involves sampling from the
 augmented \textit{complete data} posterior distribution \( p(\bm{z}, \bm{\beta}, \sigma^2 \mid \bm{y})\).\footnote{For notational brevity, we do not explicitly condition on the fixed quantities $P_i$ and $\bm{x}_i$ throughout.} For the model under consideration, this can be accomplished by iterating through the following steps:

\begin{enumerate}
    \item Sample \( \bm{z} \) from \( p(\bm{z} \mid \bm{\beta}, \sigma^2, \bm{y}) \). It is easy to show that
    {the $z_i$ are conditionally independent and each}
    full conditional is proportional to a truncated normal distribution: 
    \begin{eqnarray*}
        z_i {\mid \bm{\beta}, \sigma^2, \bm{y}}
    \sim \mathcal{TN}(\bm{x}_i^\top \bm{\beta} + P_i, \sigma^2),
    \end{eqnarray*}
    where each $z_i$ is independently drawn from a Gaussian distribution truncated to the intervals implied by (\ref{eq:choice}). Given the logarithmic link, the likelihood information restricts $z_i$ to 
    {$(-\infty, 0)$ for $y_i=0$ and to $[\log(y_i), \log(y_i+1))$ for $y_i>0$.}
    
    \item Sample \( \sigma^2 \) from \( p(\sigma^2 \mid \bm{z}, \bm{\beta}) \). The full conditional of $\sigma^2$ is proportional to an inverse gamma distribution:
     \begin{eqnarray*}
    {\sigma^2 \mid \bm{z}, \bm{\beta}}  \sim
    \mathcal{IG}\left(c_0 + \frac{n}{2}, C_0 + \frac{1}{2} {\sum_{i=1}^n} (z_i - \mathbf{x}_i^\top\bm{\beta} - P_i)^2\right).
    \end{eqnarray*}
    
    \item Sample \( \bm{\beta} \) from \( p(\bm{\beta} \mid \bm{z}, \sigma^2) \), which is proportional to a multivariate Gaussian density:
    
\[
\begin{aligned}
{\bm{\beta}  \mid \bm{z}, \sigma^2} &\sim 
\mathcal{N}(\bm{B}_N\bm{b}_N, \bm{B}_N), \\
\bm{B}_N^{-1} &= \bm{B}_0^{-1} + \frac{1}{\sigma^2}{\sum_{i=1}^n}  \bm{x}_i\bm{x}_i^\top, \quad \bm{b}_N = \frac{1}{\sigma^2} {\sum_{i=1}^n}\bm{x}_i (z_i-P_i).
\end{aligned}
\]

\end{enumerate}

\subsection{Count Size and MCMC Efficiency}
\label{sec:efficiency}

While data augmentation algorithms are typically implemented for computational convenience, they are known to result in potentially inefficient MCMC algorithms with highly autocorrelated posterior draws (\cite{duan2018scaling};\cite{zens2024ultimate}). These inefficiencies arise because the samples of \( \bm{\beta} \) and \( \bm{z} \) can be highly dependent. In count data regression, the degree of this dependency, and hence the degree of MCMC inefficiency, generally depends on the observed outcome \( y_i \).\footnote{A second major determinant of MCMC efficiency in the semi-parametric count data model considered here is the value of $\sigma^2$, as $\bm{\beta}$ and $\bm{z}$ become perfectly dependent as $\sigma^2\rightarrow0$.}

For large counts, the likelihood contributions of \( p(y_i|z_i) \) become increasingly concentrated.  As \( y_i \rightarrow \infty \), the interval \( \big[\log(y_i), \log(y_i + 1)\big) \) becomes infinitesimally small, effectively fixing \( z_i \) at a point mass in the limit. Consequently, MCMC efficiency is higher in scenarios with large counts, as \( z_i \) can be considered approximately fixed, and the dependence between \( \bm{\beta} \) and \( \bm{z} \) is minimized. This results in efficiency comparable to that of a Bayesian Gaussian linear model with fixed outcomes.

On the contrary, in the limiting case where \( y_i \rightarrow 0 \), the likelihood contributions $p(y_i|z_i)$ become minimally informative, as \( y_i = 0 \) merely implies that \( z_i < 0 \). In such cases, there is strong a posteriori dependency between $\bm{\beta}$ and $\bm{z}$
when using DA-MCMC algorithms. As the Gaussian `prior' for \( z_i \), given by (\ref{eq:latent}), then strongly determines the conditional posterior of \( z_i \), MCMC step sizes may become small.\footnote{Note that for binary regression models, all likelihood contributions exhibit behavior similar to this, leading to the typically observed inefficiencies in data augmentation algorithms for these models.} This explains the inefficient behavior observed in Figure~\ref{fig:trace_intro}.

Importantly, these considerations are not a phenomenon unique to the semi-parametric model (\ref{eq:latent})-(\ref{eq:choice}) considered here. Very similar considerations apply in the context of typical parametric regression models for overdispersed counts. For example, in the context of Poisson models with a logarithmic link where $y_i \sim \mathcal{P}(e^{z_i})$, the likelihood contributions $p(y_i|z_i)$ concentrate at a point mass at $\log(y_i)$ as $y_i \rightarrow \infty$ and degenerate at $y_i=0$, as $p(y_i|z_i) \rightarrow 1$ when $z_i \rightarrow -\infty$ if $y_i=0$.\\

\section{Marginal Data Augmentation for Bayesian Count Data Regression}

\label{sec:px-da}

\textit{Marginal data augmentation} algorithms, also referred to as \textit{parameter-expanded} data augmentation (PX-DA) algorithms, can be highly successful in accelerating DA-MCMC schemes for various latent variable regression model families (\cite{liu1999parameter}; \cite{meng1999seeking}; \cite{van2001art}; \cite{zens2024ultimate}). Such algorithms can be constructed by introducing a working parameter $\delta \sim p(\delta)$ that is unidentifiable under the observed data model but identifiable under the complete data model. Starting with the target complete data posterior $p(\bm{\beta}, \sigma^2, \bm{z} | \bm{y})$, we use the working parameter $\delta$ to define a joint posterior distribution
\begin{equation}
    p(\bm{\beta}, \sigma^2, \bm{z}, \delta | \bm{y}) = p(\bm{\beta}, \sigma^2, \bm{z} | \bm{y})~p(\delta),
\end{equation}
which we combine with an invertible and differentiable mapping $\Tilde{z}_i = \sqrt{\delta}z_i$ where $\delta=1$ results in the identity mapping. In this article, we consider a proper working prior $p(\delta) = \mathcal{IG}(d_0, D_0)$ and leave investigation of potentially more efficient but more complex algorithms relying on improper working priors for future research.

Using the working parameter $\delta$, we 
{obtain}
a set of observationally equivalent data augmentation schemes using the scale transformation $\Tilde{z}_i = \sqrt{\delta}z_i$. We will consider expansions of the form
\begin{equation}
\begin{split}
y_i&=k \Longleftrightarrow \frac{\tilde z_i}{\sqrt{\delta}}\in I_k,\quad
I_0=(-\infty,0),\\\quad I_k&=[\log(k),\log(k+1))\text{ for }k\ge1,
\\
\tilde z_i&=\sqrt{\delta} x_i^\top\beta+\tilde\varepsilon_i \quad \text{where} \quad \tilde{\varepsilon}_i \sim \mathcal{N}(0, \delta \sigma^2).
\label{eq:semiparametric_transformation}
\end{split}
\end{equation}
Importantly, we will focus on scaling $z_i$ using $\delta$ only for selected observations 
{$y_i$}.
Specifically, we focus on transforming only 
{those} $z_i$ for which $y_i=0$, while fixing $\delta=1$ for observations with $y_i>0$. This is because observations with $y_i=0$ typically cause most MCMC inefficiency, as discussed above. In addition, scaling observations with $y_i>0$ typically leads to a number of complications and potentially lower efficiency gains, as 
{will be}
discussed in more detail in Section~\ref{sec:conclusion}. Since the marginal observed-data model is unchanged by the scale expansion for any $\delta > 0$, this parameter expansion scheme is valid and can be used to construct an MCMC algorithm that directly targets the marginal posterior distribution
\begin{equation}
    p(\bm{\beta}, \sigma^2, \bm{\Tilde{z}} | \bm{y}) = \int  p(\bm{\beta}, \sigma^2, \bm{\Tilde{z}}, \delta | \bm{y}) d\delta.
\end{equation}

Sampling steps to obtain samples from $p(\bm{\Tilde{z}}, \sigma^2, \delta | \bm{y}, \bm{\beta})$ and from $p(\bm{\beta}, \delta | \bm{\Tilde{z}}, \sigma^2)$ can be constructed relatively easily, by iterating between
\begin{enumerate}
    \item updating $(\bm{\Tilde{z}}, \sigma^2, \delta) \sim p(\bm{\Tilde{z}}, \sigma^2, \delta | \bm{y}, \bm{\beta})$:
    \begin{enumerate}
    \item $z_i \sim p(z_i | y_i, \sigma^2, \bm{\beta}
    )$;
    \item $\sigma^2 \sim p(\sigma^2 | 
   {\bm{z}} ,
    \bm{\beta})$;
    \item $\delta^* \sim \mathcal{IG}(d_0, D_0)$;
    \item set $\Tilde{z}_i = z_i \sqrt{\delta^*}$ for all $i$ where $y_i=0$, otherwise $\Tilde{z}_i = z_i$;
    \end{enumerate}
    \item updating $(\bm{\beta}, \delta) \sim p(\bm{\beta}, \delta | \bm{\Tilde{z}}, \sigma^2, \bm{y})$:
    \begin{enumerate}
        \item $\delta \sim p(\delta)p(\delta | \bm{\Tilde{z}}, \sigma^2, \bm{y}) \propto p(\bm{\Tilde{z}} | \delta, \sigma^2) p(\bm{y}|\bm{\Tilde{z}}, \delta)$;
        \item set $z_i^{\text{new}} = \Tilde{z}_i / \sqrt{\delta}$ for all $i$ where $y_i=0$, otherwise ${z}^{\text{new}}_i = z_i$;
        \item $\bm{\beta} \sim p(\bm{\beta} | \bm{z}^{\text{new}}, \sigma^2)$.
    \end{enumerate}
\end{enumerate}
As $\delta$ is sampled in both steps, iterating between sampling from these two joint distributions allows investigating the posterior in $(\bm{\Tilde{z}}, \bm{\beta}, \sigma^2)$ marginalized over $\delta$, which typically increases MCMC efficiency, due to the posterior of $\bm{\Tilde{z}}$ being more diffuse than the posterior of $\bm{{z}}$, see \cite{liu1999parameter}, \cite{meng1999seeking}, \cite{van2001art}, among others. Compared to the plain DA algorithm outlined in the previous section, this strategy leads to significant gains in MCMC efficiency at negligible additional computational costs.

Conceptually, such algorithms can be seen as extending the standard sampling schemes to include a `sandwiched' in-between step, leading to algorithms of the form 
\begin{enumerate}
    \item sample $(\bm{z}, \sigma^2) \sim p(\bm{z}, \sigma^2| \bm{\beta}, y_i)$;
    \item move $\bm{z} \rightarrow \bm{z}^{\text{new}}$;
    \item sample $\bm{\beta} \sim p(\bm{\beta}| \bm{z}^{\text{new}}, \sigma^2)$;
\end{enumerate}
where the move from $\bm{z} \rightarrow \bm{z}^{\text{new}}$ based on $z_i^{\text{new}} = \sqrt{\frac{\delta^*}{\delta}}z_i$ is used to increase the step size of the MCMC algorithm and decrease the correlation between $\bm{\beta}$ and the latent outcomes $\bm{z}$. 

\paragraph{Implementation details}

Most of the necessary updating steps are straightforward to derive. For instance, all updates required to sample \((\bm{\Tilde{z}}, \sigma^2, \delta) \sim p(\bm{\Tilde{z}}, \sigma^2, \delta \mid \bm{y}, \bm{\beta})\) are standard and easy to implement. Similarly, once \(z_i^{\text{new}}\) has been obtained, sampling step (2c), which involves sampling from \(p(\bm{\beta} \mid \bm{z}^{\text{new}}, \sigma^2)\), is straightforward. The necessary steps follow along the lines of the algorithm given in Section~\ref{sec:model}. 

The only major deviation from the described plain data augmentation scheme arises due to sampling {the working parameter $\delta$}
from \(p(\delta \mid \bm{\Tilde{z}}, \sigma^2, \bm{y}) \propto p(\delta)p(\bm{\Tilde{z}} \mid \delta,  \sigma^2) \, p(\bm{y} \mid \bm{\Tilde{z}}, \delta)\). As we are expanding only observations with $y_i=0$, the likelihood contribution \(p(\bm{y} \mid \bm{\Tilde{z}}, \delta)\) is irrelevant for this step, since scaling does not affect the observed data model in these cases and \(p(\bm{y} \mid \bm{\Tilde{z}}, \delta)=1\) for all relevant values of $\bm{\Tilde{z}}$.

To derive the density $p(\bm{\Tilde{z}} \mid \delta, \sigma^2)$, we proceed as follows. First, we derive the posterior $p(\bm{\beta}|\sigma^2, \delta, \Tilde{\bm{z}})$, which is a multivariate Gaussian
{distribution}. Second, we exploit Bayes' theorem and evaluate the following ratio at $\bm{\beta}=0$:
\begin{equation}
    p(\bm{\Tilde{z}} \mid \delta, \sigma^2) \propto \frac{p(\bm{\beta})\prod_{i:y_i=0} p(\Tilde{z}_i| \bm{\beta}, \sigma^2, \delta
   ) 
    \prod_{i:y_i>0}  p(\Tilde{z}_i| \bm{\beta}, \sigma^2)}{p(\bm{\beta}|\sigma^2, \delta, \Tilde{\bm{z}})}.
\end{equation}
Combining the resulting closed-form likelihood term  $p(\bm{\Tilde{z}}\mid\delta, \sigma^2)$ with the working prior $p(\delta)=\mathcal{IG}(d_0, D_0)$ yields
\begin{equation}
\label{eq:delta_density}
p(\delta \mid \delta^*, \bm{\Tilde{z}}, \sigma^2) \propto \left(\frac{1}{\delta}\right)^{d_I+1} e^{-D_I/\delta} e^{B_I/\sqrt{\delta}},
\end{equation}
with 
\begin{equation}
    \begin{aligned}
      d_I &= d_0 + \frac{n_0}{2},\\
      D_I &= D_0 + \frac{\delta^*}{2\sigma^2} \sum_{i:y_i=0}z_i^2 - \frac{\delta^*}{2} \mathbf{m}_0^\top \mathbf{B}_N \mathbf{m}_0 ,\\
      B_I &= \sqrt{\delta^*} \left(%
      {\frac{1}{\sigma^2}\sum_{i:y_{i}=0} z_{i} P_{i}}
      +  \mathbf{m}_0^\top \mathbf{B}_N (\mathbf{m}_1 - \mathbf{m}_2) \right),
    \end{aligned}
\end{equation}
where $n_0 = \sum_i\mathbbm{1}(y_{i}=0)$
{is the number of observed zero counts}, 
$\mathbf{m}_0 = {\sigma^{-2}}
\sum_{i:y_{i}=0} \mathbf{x}_i z_{i}$, $\mathbf{m}_1 = \sigma^{-2}\sum_{i:y_{i}>0} \mathbf{x}_i z_{i}$, and $\mathbf{m}_2 = \sigma^{-2}\sum_i \mathbf{x}_i P_{i}$. The conditional posterior for \(\delta\) hence takes the form of a density similar, but not equivalent to an inverse gamma density.\footnote{Note that although \(\delta\) is unidentifiable from the observed data, its conditional posterior moments depend on the latent data \(z_i\) because the re-parameterisation \(\tilde z_i = \sqrt{\delta}z_i\) `injects' \(\delta\) into the augmented joint likelihood $p(\bm{y}, \tilde{\bm{z}}|\cdot)$.} This nuisance is the result of expanding only observations where $y_i=0$ and the presence of an offset term $P_i$. Note that an inverse gamma posterior is recovered, in particular, when all observations are expanded jointly. An efficient algorithm to sample from density (\ref{eq:delta_density}) is described in the supplementary material of \cite{zens2024ultimate}, where more details regarding the derivation can be found as well. For completeness, we replicate the necessary sampling steps in Appendix~\ref{appA}.

\section{Illustration Using Artificial Data}

\label{sec:simulations}

To evaluate the sampling efficiency of the proposed PX-DA scheme, we consider several experiments based on artificial data. In all cases, we generate $n=1{,}000$ observations according to $y_i = \lfloor e^{z_i} \rfloor$, with the latent Gaussian regression model specified as
\begin{equation*}
z_i = \beta_0 + \bm{x}_i^\top \bm{\beta} + \varepsilon_i,\quad \varepsilon_i \sim \mathcal{N}(0, \sigma^2).
\end{equation*}
We explore a variety of simulation settings. The intercept term $\beta_0$ is varied over the interval $[-4,4]$, ranging from cases where nearly all {$y_i$ are equal to zero}
and MCMC efficiency is expected to be low, to situations where no zeros are observed and near-optimal MCMC efficiency is achieved. In addition, we consider $\sigma^2 \in \{0.05, 0.5\}$ to assess different levels of informativeness of the latent process.  Regarding the covariates, we consider an intercept-only case ($\bm{\beta}=\bm{0}$) and two settings with $p=4$ covariates, each simulated independently from $\mathcal{N}(0,1)$. In the first setting, the predictors are relatively informative with
{$\bm{\beta}=(1,-1,0.5,-0.5)^\top$}
while in the second setting they are less informative with 
{$\bm{\beta}=(0.1,-0.1,0.05,-0.05)^\top$.}

In terms of priors, we assume $\beta_j \sim \mathcal{N}(0,100)$ for both $\beta_0$ and the elements of $\bm{\beta}$, and $\sigma^2 \sim \mathcal{IG}(5,1)$. We collect $M=20{,}000$ posterior samples after an initial burn-in period of $5{,}000$ iterations. The primary measure of sampling inefficiency that we consider is the inefficiency factor, defined as $M/\hat{\eta}$, where $\hat{\eta}$ is an estimate of the effective sample size obtained under a given algorithm.\footnote{For each parameter, the effective sample size is derived based on an estimate of the spectral density of the posterior sample chain at frequency zero, using the functionality provided in the \texttt{R} package \texttt{coda} (\cite{codaR}).} Each simulation setting is replicated 50 times, and we report the average results across these replicate data sets below.

\begin{figure}
    \centering
    \includegraphics[width=\linewidth]{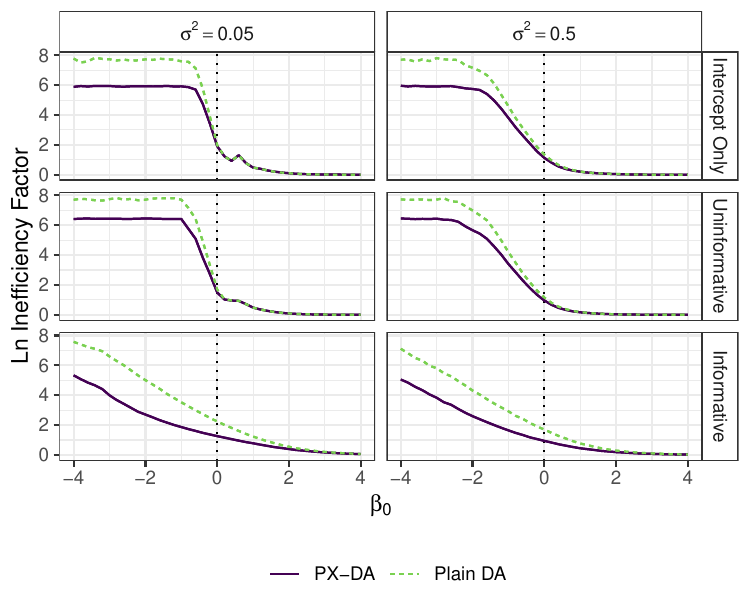}
    \caption{Inefficiency factors (logarithmic scale) obtained under plain DA and PX-DA algorithms for the considered simulation scenarios (panels) and values of intercept $\beta_0$ ($x$-axis). Results are averages across all parameter blocks and 50 replicate data sets.}
    \label{fig:ie_all}
\end{figure}

Fig.~\ref{fig:ie_all} summarizes the main findings of this simulation study. It shows inefficiency factors averaged over all parameters for both the plain DA algorithm and the PX-DA algorithm. It is evident that the PX-DA scheme markedly outperforms the plain DA scheme in every setting. Averaged across all settings and parameters, the PX-DA scheme leads to a decrease in inefficiency factors by about 79\%, from $M/\hat{\eta} \approx 580$ in the plain DA algorithm, to $M/\hat{\eta} \approx 124$ in the PX-DA algorithm. In the most extreme cases, the PX-DA scheme is either similarly efficient (difference less than 1\%) or significantly more efficient (up to more than 90\% improvement) compared to the plain DA scheme. 

Across all simulation scenarios, the PX-DA algorithm shows greater relative improvement when $\beta_0$ is strongly negative, when $\sigma^2$ is small, and in cases where the covariates are highly informative. This occurs because, all else being equal, a latent model with a coefficient of determination near one results in a stronger dependence between the latent outcome and the regressors, reducing sampling efficiency. Similarly, smaller values of $\sigma^2$ increase the dependence between $z_i$ and $\bm{\beta}$, further lowering sampling efficiency. In these scenarios, the benefits of the PX-DA algorithm are particularly pronounced.

In Appendix~\ref{appB}, additional simulation results are provided, broken down by (blocks of) parameters $\beta_0$, $\bm{\beta}$, and $\sigma^2$ (Figs.~\ref{fig:ie_alpha}-\ref{fig:ie_sigma2}). The proposed PX-DA algorithm dominates the plain DA algorithm in all blocks. The gains are especially pronounced
for the intercept term $\beta_0$, followed by the gains in efficiency for the coefficients $\bm{\beta}$. The efficiency improvement is least substantial for the error variance $\sigma^2$. 

\section{Application to Latent Factor Models for Subnational Mortality Counts}

\label{sec:application}

One of the primary aims of statistical demography is to develop effective models for human mortality, fertility, and migration patterns, providing essential insights for population projections and regional planning exercises. Demographic models can be used to track and predict demographic trends, potentially across multiple populations, and to impute data where reliable observations are missing -- the latter issue being particularly relevant in low-income countries and in subnational analyses. 

In recent decades, Bayesian approaches have emerged as a dynamic and growing subfield within this domain (\cite{raftery2012bayesian}; \cite{raftery2014bayesian}; \cite{bijak2016bayesian}; \cite{wisniowski2015bayesian}; \cite{alexander2017flexible}; \cite{zens2024flexible}). In a pioneering paper \cite{czado2005bayesian}, Claudia Czado and her co-authors explored fully Bayesian factor models for demographic analysis.  Similar approaches have been used by \cite{wisniowski2015bayesian} and \cite{zens2024flexible}, among others.

When analyzing demographic patterns by age group, low-rank approximation techniques -- such as singular value decompositions or parametric factor models -- have demonstrated particular effectiveness (\cite{lee1992modeling}; \cite{wisniowski2015bayesian}; \cite{alexander2017flexible}; \cite{clark2019general}; \cite{zens2024flexible}). This is due to the nature of demographic phenomena, which are often closely related to life course events or biological processes that generate common variation across diverse populations. For instance, increasing mortality rates at older ages represent a `universal' pattern regardless of the specific population under study. 

However, estimation of such low-rank models in combination with count data likelihoods remains challenging, especially in settings with a relatively high prevalence of zero counts—a common occurrence when disaggregating data to granular subnational levels. This motivates our evaluation of the proposed PX-DA scheme in this context. We implement and estimate a semi-parametric Bayesian latent factor model for count and rate data and apply it to subnational mortality data from Austria. Importantly, this allows us to evaluate the applicability of the proposed methodology in a real-world application that goes beyond classical regression modeling.

\subsection{Model and Prior Specification}

We consider mortality counts $y_{ia}$ recorded for a subpopulation $i = 1, \dots, K$ and age group $a = 1, \dots, A$. 
{The observations are collected in the $K\times A$ data array $\bm{y}$.}
We assume that the counts $y_{ia} = \lfloor \exp(z_{ia}) \rfloor$ can be well represented using a latent factor model on the latent utilities $z_{ia}$: 
\begin{equation}
\label{eq:factormodel}
    z_{ia} = \mu_i + {\sum_{q=1}^Q} f_{aq}\lambda_{iq} + P_{ia} + \varepsilon_{ia}, \qquad
    {\varepsilon_{ia}}
    \sim \mathcal{N}(0, \sigma^2_i),
\end{equation}
where $f_{aq}$ $(q=1, \dots, Q)$ are latent factors that capture common variation in the age dimension, $\lambda_{iq}$ are subpopulation-specific factor loadings, $\mu_i$ is a subpopulation-specific intercept term, and {$\varepsilon_{ia}$}
is a Gaussian random effect with subpopulation-specific variance $\sigma^2_i$, accounting for overdispersion and stochastic variation. $P_{ia}$ denotes a known offset term, representing the log-transformed `population-at-risk'. This allows the model to be interpreted in terms of mortality \textit{rates} and provides some standardization between subpopulations.
Due to the regularities of demographic processes, a number of factors $Q \ll K$
can be expected to allow for an accurate representation of the $K$ observed age patterns. Hence, (\ref{eq:factormodel}) provides a parsimonious modeling and denoising framework for granular, subnational demographic processes.

Simple, weakly informative and conditionally conjugate $\mu_i \sim \mathcal{N}(0, M_0)$, $\lambda_{iq} \sim \mathcal{N}(0,L_0)$ and $\sigma^2_i \sim \mathcal{IG}(c_0, C_0)$ priors are specified for the factor loadings and the subpopulation-specific error variances, where we used $c_0=2.5$, $C_0 = 1.5$ and $M_0 = L_0 = 100$ as hyperparameters. Extensions to more complex hierarchical prior setups as in \cite{alexander2017flexible} or \cite{zens2024flexible} are straightforward to accommodate. In terms of factors, we use independent $f_{aq} \sim \mathcal{N}(0,1)$ priors for $q=1, \dots, Q$ and $a=1, \dots, A$ to enhance scale identification during MCMC.%

\subsection{Posterior Simulation}
\label{sec:update_f}
Conditional on all other parameters, the latent factors $f_{aq}$ for
{$a=1, \ldots,A$ and}
$q=1, \dots, Q$ can be updated as follows. First, compute the working observations $z^*_{ia} = z_{ia} - \mu_i - P_{ia}$ for all $i = 1, \dots, K$ and $a = 1, \dots, A$. For each $a = 1, \dots, A$, 
the conditional posterior distribution of the 
{age-specific factor $\bm{f}_a=(f_{a1}, \ldots,f_{aQ})^\top$} 
is proportional to a multivariate Gaussian density $\bm{f}_a {|\cdot} \sim \mathcal{N}_{Q}(\bm{f}_{N,a}, \bm{F}_{N})$ with posterior moments 
\begin{equation*}
    \mathbf{F}_{N} = \left(\bm{I}_{Q} + \mathbf{\Lambda}^\top\bm{\Sigma}^{-1}\mathbf{\Lambda}\right)^{-1} ,\quad \quad \mathbf{f}_{N,a} = \mathbf{F}_N\mathbf{\Lambda}^\top\bm{\Sigma}^{-1}\bm{z}^*_a,
\end{equation*}
where $\mathbf{\Lambda}$ denotes the $K \times Q$ matrix of
{all factor loadings $\lambda_{iq}$},  
$\bm{\Sigma} = \text{diag}{(\sigma^2_1, \ldots, \sigma_K^2)}$
and $\bm{z}^*_a = (z^*_{1a}, \dots, z^*_{Ka})^\top$. Conditional on the latent factors $f_{aq}$, the posterior simulation scheme follows the PX-DA scheme outlined in Section~\ref{sec:px-da} with obvious adaptations. For completeness, the full PX-DA scheme for the latent factor model can be found in App.~\ref{appC}.

\subsection{Data}

We estimate the factor model (\ref{eq:factormodel}) using data on age- and sex-specific subnational mortality patterns in Austria in 2023. The data stem from the national statistical office of Austria (\textit{Statistik Austria}) and are publicly available.\footnote{\url{https://www.statistik.at/en/databases/statcube-statistical-database/free-access}} The data cover mortality counts for $A=20$ five-year age groups in 116 subnational units (corresponding to 93 political districts outside Vienna and the 23 districts of Vienna) for males and females, resulting in $K=232$ subpopulations. As is typical for subnational mortality counts in high-income countries, a large proportion (23.8\%) of the $N=4{,}640$ observations are zero, due to relatively small populations and low mortality in younger age groups. In addition to the mortality counts, we collect data on age- and sex-specific population counts for the same subnational units on January 1, 2023. The natural logarithm of these population counts is added as an offset $P_{ia}$ in the model.

\begin{figure}
    \centering
\begin{subfigure}[b]{0.49\textwidth}
        \centering
        \includegraphics[width=\linewidth]{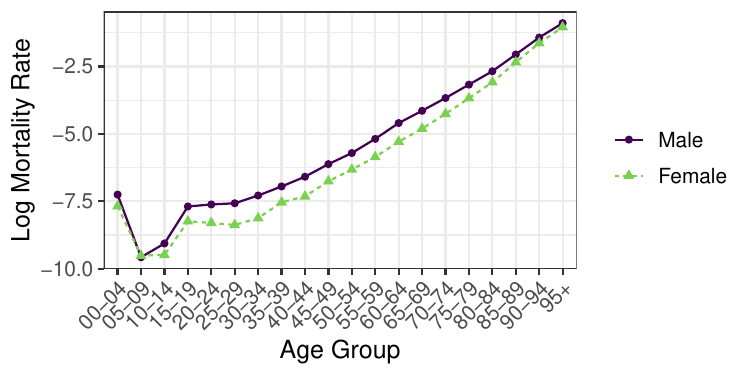}
        \caption{Aggregate Mortality}
    \end{subfigure}
    \hfill
    \begin{subfigure}[b]{0.49\textwidth}
        \centering
        \includegraphics[width=\linewidth]{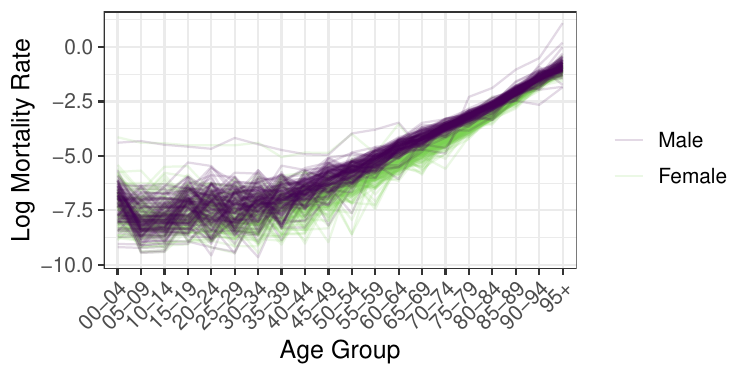}
        \caption{Subnational Mortality}
    \end{subfigure}
        \caption{Aggregate (a) and district-specific (b) log-transformed mortality rates ($y$-axis) by sex (color) and five-year age groups ($x$-axis) for Austria in 2023. For visualization purposes, observed mortality counts and population counts are offset by 0.5 before log-transformation to avoid undefined logarithms of zero values.}
    \label{fig:mortality}
\end{figure}

The data are visualized in Fig.~\ref{fig:mortality}. Panel (a) shows log mortality rates disaggregated by sex, based on countrywide aggregate mortality and population counts. The patterns are very typical, including male mortality exceeding female mortality, rising mortality at older ages, slightly elevated mortality risk in the youngest age group, mostly due to infant mortality, and the so-called \textit{accident hump} at ages 15-19, which is particularly pronounced among males and is typically attributed to a rise in behavioral causes of death in this age group. Panel (b) shows all subnational log mortality curves superimposed. The striking similarities in the empirical patterns across all subpopulations illustrate why low-rank factor models can be highly successful in representing high-dimensional demographic data. 

\subsection{Model Selection}

To select an appropriate number of latent factors \(Q\), we compared several methods and criteria. These included a heuristic examination of the step-wise improvement of in-sample log predictive density scores, pseudo-out-of-sample log predictive scores as in \cite{czado2009predictive}
in combination with ten-fold cross-validation, and other criteria that focus on approximating out-of-sample predictive accuracy, such as the WAIC (as used in \cite{kowal2020simultaneous}), or an estimate of the expected log predictive density based on Pareto smoothed importance sampling (as proposed in \cite{vehtari2017practical}).\footnote{Conveniently, the integrated likelihood $\int p(y|z)p(z|\beta, \sigma^2) dz$ is available in closed-form in the model under consideration, facilitating computation of log predictive scores for model selection. The relevant quantity is given by 

\begin{equation}
p(y_{ia}=k)=
\begin{cases}
\Phi\left(\dfrac{-\mu_{ia}}{\sigma_i}\right), & k=0, \\[6pt]
\Phi\left(\dfrac{\log(k+1)-\mu_{ia}}{\sigma_i}\right)
-
\Phi\left(\dfrac{\log(k)-\mu_{ia}}{\sigma_i}\right), & k\ge 1,
\end{cases}
\end{equation}

where $\Phi(\cdot)$ denotes the cdf of the standard Gaussian distribution and $\mu_{ia} = \mu_i + \sum_q f_{aq}\lambda_{iq} + P_{ia}$ is the conditional mean of the Gaussian model for $z_{ia}$
{defined in (\ref{eq:factormodel})}.
} All of these methods point towards models with $Q$ relatively large. A variant of the \textit{joint-likelihood-based information criterion} (JIC) -- as introduced in \cite{chen2022determining} for choosing the number of latent factors in generalized linear factor models -- based on in-sample log predictive density scores instead clearly favors a model with a single latent factor.\footnote{The original version of the JIC for a model with $Q$ latent factors is given by $\text{JIC}_Q = -2\bm{l}_Q + Q\max(K,A)\log\{\min(K,A)\}$ where $\bm{l}_Q$ denotes the likelihood obtained at the maximum likelihood parameter estimates.}%

Overall, these methods did not converge on a single answer, underscoring the significant challenges of model selection in latent factor models -- challenges that are even more pronounced in non-Gaussian contexts. Based on additional simulation experiments evaluating the performance of these criteria when $Q$ is known, and in the interest of parsimony, we focus on the model favored by the JIC, as this criterion most often led to the correct model choice in our experiments. Similar findings motivated the use of the JIC for model selection in non-Gaussian factor models in \cite{mauri2024factor}. Further evidence that a model with 
{a single factor}
is useful comes from examining a singular value decomposition of the log-transformed mortality rates, where the leading singular triplet explains more than 96\% of the variance in the log-transformed data. Moreover, the single-factor case is consistent with the most classical mortality model in statistical demography, the Lee-Carter model (\cite{lee1992modeling}). A detailed investigation of the performance of different model selection criteria in latent variable factor models is left for future research.

\subsection{Results}

\begin{figure}[t]
    \centering
    \begin{subfigure}[b]{0.32\textwidth}
        \centering
        \includegraphics[width=\textwidth]{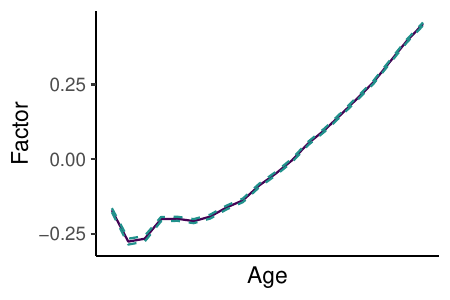}
        \caption{Age factor $f_a$.}
    \end{subfigure}
    \hfill
    \begin{subfigure}[b]{0.32\textwidth}
        \centering
        \includegraphics[width=\textwidth]{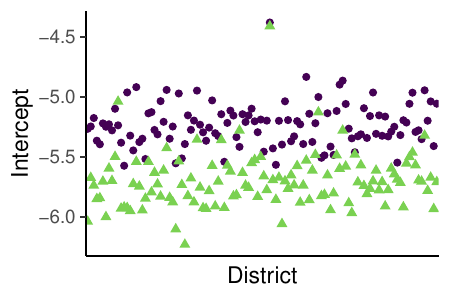}
        \caption{Intercept $\mu_i$.}
    \end{subfigure}
       \hfill
    \begin{subfigure}[b]{0.32\textwidth}
        \centering
        \includegraphics[width=\textwidth]{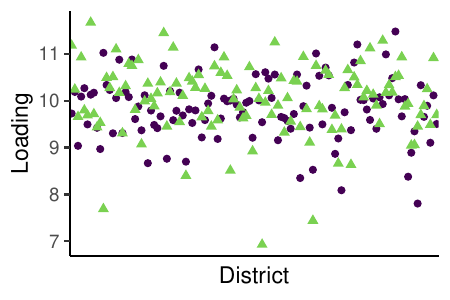}
        \caption{Loadings $\lambda_i$.}
    \end{subfigure}
    \caption{Estimated factor and loadings from the mortality model {with $Q=1$}. The left panel shows the estimated factor with 95\% credible intervals. The middle panel displays subpopulation-specific intercept terms. The right panel shows loadings on the age-varying factor. All point estimates are posterior means. Colors and shapes refer to male (purple circle) and female (green triangle) populations.}
    \label{fig:factors_loadings}
\end{figure}

Fig.~\ref{fig:factors_loadings} displays the estimated {single} factor (left panel), estimates of the subpopulation-specific intercept terms (middle panel), and the estimated loadings for the latent factor (right panel). To address scale identifiability in the latent factor model, posterior samples of both factors and loadings from the MCMC chain were rescaled ex post prior to calculating summary statistics for this visualization.
All point estimates shown are posterior means. The factor plot (left panel) includes 95\% credible intervals, demonstrating that the factor is estimated with high precision. The factor clearly captures the empirical mortality patterns discussed earlier. The intercept terms reveal average mortality differences between male (purple) and female (green) populations, while the factor loadings indicate that the estimated factor applies to all subpopulations, without exception. Note that gender information was not provided to the model during estimation—the model detected these subpopulation patterns without any knowledge of male/female status, with gender labels applied here solely for interpreting and visualizing the results.
Refined versions of the model could incorporate hierarchical structures on the loadings and intercept terms to enable further information sharing between subpopulations; see \cite{alexander2017flexible} and \cite{zens2024flexible} for examples.

\begin{figure}[t]
    \centering
    \includegraphics[width=\linewidth]{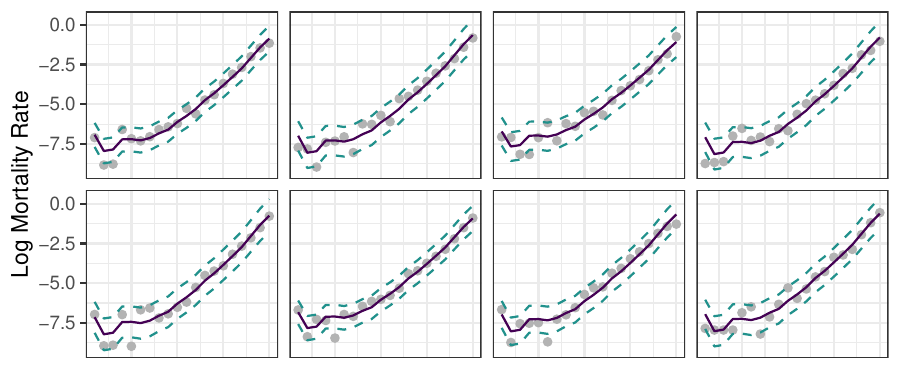}
    \caption{Data points $\log y_{ia}-P_{ia}$ (grey points), as well as posterior means (purple solid) and 95\% credible intervals (blue dashed) of the posterior predictive densities of $z_{ia}-P_{ia}$ for the $A=20$ age groups ($x$-axis, suppressed) and eight exemplary subpopulations; based on the model with $Q=1$. For visualization purposes, observed mortality counts, population counts, and posterior predictive draws are each offset by 0.5 before log-transformation to avoid undefined logarithms of zero values.}
    \label{fig:fitted_values}
\end{figure}

Fig.~\ref{fig:fitted_values} compares observed log mortality rates against the posterior predictive density of $z_{ia}-P_{ia}$ for eight representative subpopulations to illustrate model fit. The figure presents posterior means and 95\% {predictive}
intervals derived from samples of the predictive density. Overall, the model demonstrates satisfactory fit, particularly considering its parsimony with only a single factor.

Regarding estimation efficiency, we first compare MCMC efficiency of the fitted values $\mu_i + \sum_q f_{aq}\lambda_{iq}$ for each $i = 1, \dots, K$ and $a = 1, \dots, A$. The PX-DA algorithm yields an average 16.6\% point-wise improvement in inefficiency factors and reduces average inefficiency factors from approximately $1.98$ to $1.41$. For certain coordinates, efficiency improvements reach up to 80\%. For the error variances $\sigma_i^2$, the PX-DA algorithm provides an average 3.5\% point-wise improvement in inefficiency factors.

\section{Discussion and Conclusion}\label{sec:conclusion}

This paper introduces a marginal data augmentation approach for efficient Bayesian estimation of count data regression models. Our proposed method addresses a critical challenge in Bayesian count data analysis: the poor mixing of common data augmentation algorithms in data sets with a high proportion of zero outcomes.
By introducing a working parameter to scale latent variables specifically for such zero-count observations, our approach effectively reduces the strong posterior dependencies that typically impair standard data augmentation schemes. The performance gains of our proposed PX-DA algorithm can be substantial, as evidenced by extensive simulation studies. Our application to subnational mortality counts in Austria further demonstrates the broader applicability of the proposed methodology.

It is worth noting that we deliberately focused on boosting only observations where $y_i = 0$, as these cases typically contribute most to posterior autocorrelation. Extending the boosting scheme to include observations where $y_i > 0$ introduces technical complications that likely outweigh potential benefits. Specifically, incorporating observations with positive counts would necessitate truncating the posterior distribution of $\delta$  to respect the likelihood information, with larger counts requiring increasingly smaller scaling adjustments. For instance, in the presence of even only a single very large count, the posterior distribution of $\delta$ would concentrate at $\delta^* \approx \delta$, effectively nullifying the boosting effect for all observations. While it may be possible to identify a `sweet spot' — for example, by boosting only observations where \( y_i = 0 \) and \( y_i = 1 \), possibly with two separate scale adjustments — the efficiency gains of such an approach may not justify the additional computational and theoretical complexity. We leave investigating such more complex PX-DA schemes for future research.

Several additional promising directions for future research emerge from this work. Further investigation into marginal data augmentation schemes for count data regression models could yield further efficiency gains, particularly for data sets consisting almost exclusively of small and zero counts, where even our current method faces limitations. For instance, beyond the scaling step, a second 'sandwiched' translation step as suggested by \cite{zens2024ultimate} could potentially provide additional efficiency gains. 

In addition, the role of link functions in DA-MCMC and marginal data augmentation schemes remains underexplored. For example, degenerate likelihoods leading to low MCMC efficiency for $y_i=0$ (see Subsection~\ref{sec:efficiency}) are an issue for Poisson models with logarithmic link functions, but less so for Poisson models with square root links. More generally, developing efficient MCMC solutions for large-sample count data sets, where updating the $z_i$ becomes a dominant constraint, would be valuable. Finally, the parameter expansion ideas presented here could be extended to accelerate variational or expectation-maximization algorithms, an approach that has proven effective in other contexts such as probit regression (\cite{liu1998parameter}).

\backmatter


\clearpage

\begin{appendices}

\section{Details on Sampling $\delta$}\label{appA}

The  following  resampling technique is used 
{in \cite{zens2024ultimate} to sample $ \delta$} from 
\begin{eqnarray*}
p( \delta|  \cdot) \propto  \left( \frac{1}{\delta} \right) ^{d_I+1} e ^{- D_I/\delta}  e ^{ B_I /\sqrt{\delta}}. 
\end{eqnarray*}
First, choose an  `auxiliary  prior'  $\pi(\delta)$ for resampling such that the mode  and the curvature of $\pi(\delta)$ coincide with the  mode $\delta_M$ and the curvature $I_p$  of the posterior $p ( \delta | \cdot )$ which are  given by:
\begin{eqnarray*}
\delta_M= \frac{16 D_I^2}{\left[B_I+\sqrt{B_I^2+16 D_I(d_I+1)}\right]^2},
\qquad I_p=  -   \frac{\sqrt{B_I^2+16 D_I(d_I+1)}}{4 \cdot (\delta_M) ^{\frac{5}{2}}}.
\end{eqnarray*}
The factorization of the posterior suggests an inverse gamma distribution $\delta \sim \mathcal{G}^{-1}(d_I^\star ,D_I^\star)$ as auxiliary prior $\pi(\delta)$. The mode $\delta_{IG}$ and curvature $I_{IG}$ of the pdf are given by:
   \begin{eqnarray*}
&& \displaystyle \delta_{IG} = \frac{D_I ^\star}{d_I ^\star +1},  \qquad  I_{IG} =
-   \frac{(d_I ^\star +1)^3}{(D_I ^\star)^2}.
\end{eqnarray*}
Matching the mode, i.e. $\delta_{IG}=\delta_{M} $,  and the curvature, i.e. $I_{IG}=I_p$, to the posterior yields the following optimal choice for the parameters $(d_I^\star ,D_I^\star)$:
 \begin{eqnarray*}
d_I ^\star  = - I_p \cdot(\delta_{M})^2 -1 ,  \quad  D_I ^\star = \delta_{M} (d_I ^\star +1).
\end{eqnarray*}
 Resampling then works as follows. $L$ draws  $\delta_l \sim \pi(\delta)$, $l=1,\dots,L$,   from
 the  auxiliary prior  are resampled using  weights proportional to the `auxiliary  likelihood'  $\ell(\delta)= p( \delta|\cdot)/\pi(\delta)$, given by:
 \begin{eqnarray*}
  \log \ell(\delta) &\propto& \log p ( \delta |  \cdot) - \log \pi(\delta) \\
& \propto&
 -(d_I+1) \log \delta - \frac{D_I}{\delta}  + \frac{B_I}{\sqrt{\delta}}  - \log \pi(\delta).
\end{eqnarray*}
The desired draw from $p( \delta |\cdot ) $ is given by
 $\delta_{l^*}  $, where $ l ^\star   \sim \text{Multinomial}(1; w_1, \dots, w_L)$ and the weights $w_l \propto \ell(\delta)$   are normalized to 1. The auxiliary likelihood $\ell(\delta)$ is expected to be rather flat over the support of  $\pi(\delta)$. Hence, $(w_1, \dots, w_L)$ is expected to be close to a uniform distribution and $L$ can be small ($L=5$ or $L=10$ can be enough).

\newpage

\section{Additional Simulation Results}\label{appB}

\begin{figure}[!h]
    \centering
    \includegraphics[width=0.65\linewidth]{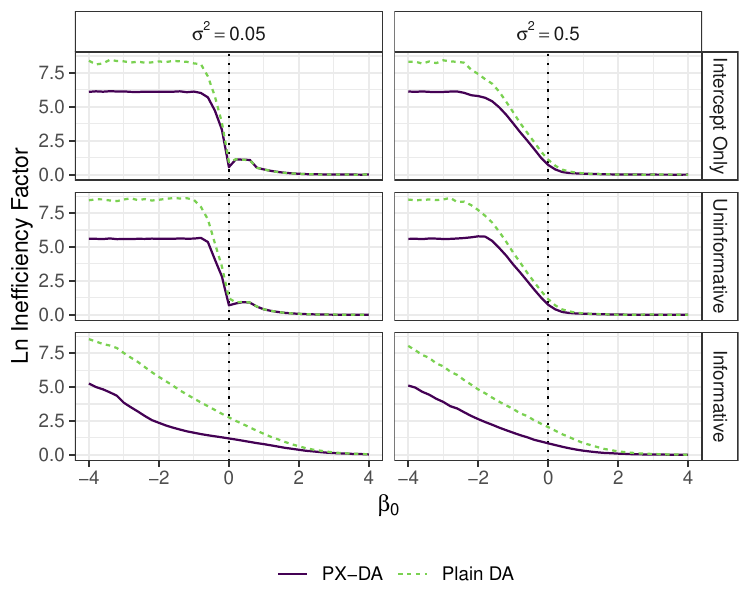}
    \caption{Inefficiency factors (logarithmic scale) for $\beta_0$ obtained under plain DA and PX-DA for various simulation scenarios and values of intercept $\beta_0$. Results are averages across 50 replicate data sets. Refer to 
    {Section~\ref{sec:simulations}}
    for details on the simulation settings.}
    \label{fig:ie_alpha}
\end{figure}

\begin{figure}[!h]
    \centering
    \includegraphics[width=0.65\linewidth]{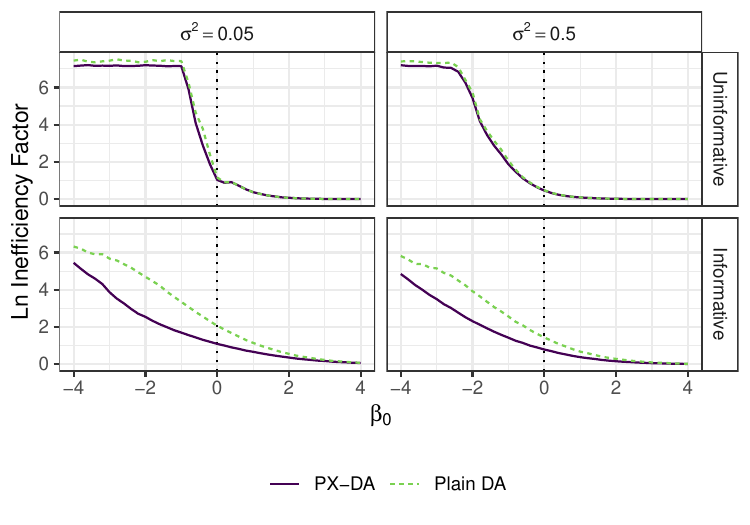}
    \caption{Inefficiency factors (logarithmic scale)  for $\bm{\beta}$ obtained under plain DA and PX-DA for various simulation scenarios and values of intercept $\beta_0$. Results are averages across all elements of $\bm{\beta}$ and 50 replicate data sets. Refer to 
{Section~\ref{sec:simulations}}
    for details on the simulation settings.}
    \label{fig:ie_beta}
\end{figure}

\begin{figure}[!h]
    \centering
    \includegraphics[width=0.7\linewidth]{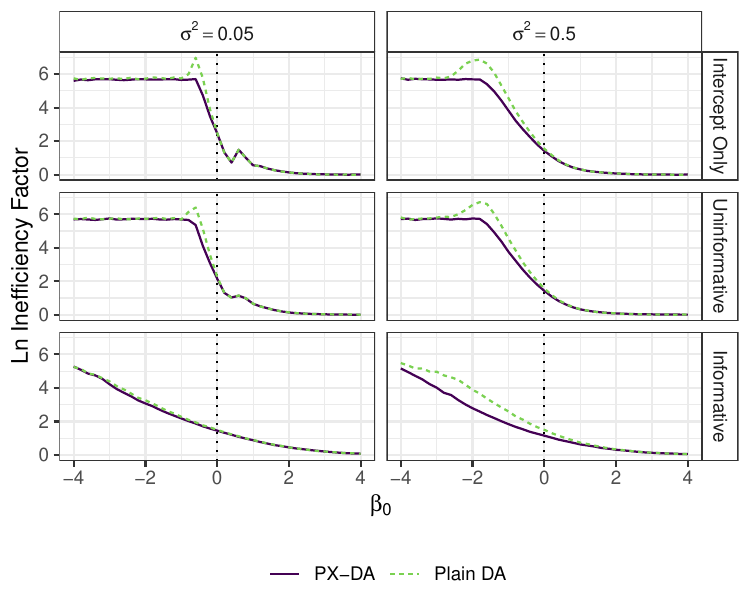}
    \caption{Inefficiency factors (logarithmic scale)  for $\sigma^2$ obtained under plain DA and PX-DA for various simulation scenarios and values of intercept $\beta_0$. Results are averages across 50 replicate data sets. Refer to 
    {Section~\ref{sec:simulations}}
    for details on the simulation settings.}
    \label{fig:ie_sigma2}
\end{figure}

\section{PX-DA for Latent Factor Models}\label{appC}

Given $y_{ia} = \lfloor\exp(z_{ia})\rfloor$, simulating from the joint posterior density of the latent data $\bm{z}$, the parameters $(\mu_i, \lambda_{iq}, \sigma^2_i)$ and the latent factors $\bm{F}$ can be based on a slightly modified version of the Gibbs scheme outlined in Section~\ref{sec:px-da}.

\begin{enumerate}
\item Update the 
{rows of the $A\times Q$}
latent factor matrix $\bm{F} = {\left(\bm{f}_1, \dots, \bm{f}_A \right) ^\top}$ as described in Section~\ref{sec:update_f}.

\item Conditional on the factors $\bm{F}$, the subpopulation-specific parameters are independent. For each {subpopulation} $i = 1, \dots, K$:

\begin{enumerate}
    \item Sample {$\bm{z}_i=( z_{i1}, \ldots,  z_{iA})^\top$}
    from \( p(\bm{z}_i \mid \mu_i, \bm{\lambda}_i, \sigma_i^2, \bm{y}_i, \bm{F}) \),
    {where $\bm{y}_i$ is the $i$th row of the data matrix $\bm{y}$ and $\bm{\lambda}_i=(\lambda_{i1},
    \ldots, \lambda_{iQ})$ are the factor loadings relevant for subpopulation $i$. Due to conditional independence,  
    the full conditional of each $z_{ia}$}    
    is proportional to a truncated normal distribution: 
    \begin{equation*}
    z_{ia} {|\cdot}  \sim \mathcal{TN}( \mu_i + \sum_q f_{aq}\lambda_{iq} + P_{ia}, \sigma_i^2), 
      \end{equation*}
  where
  $z_{ia}$ is 
  truncated to the interval 
  { $(-\infty,0)$ in case $y_{ia}=0$ and to $[\log(y_{ia}), \log(y_{ia}+1))$ otherwise.}
    
    \item Sample \( \sigma_i^2 \) from \( p(\sigma_i^2 \mid \bm{z}_i, \mu_i, \bm{F}, \bm{\lambda}_i) \). The full conditional of $\sigma_i^2$ is proportional to an inverse gamma distribution:
   \begin{equation*}
    {\sigma^2 _i} {|\cdot} 
    \sim  \mathcal{IG}\left(c_0 + \frac{A}{2}, C_0 + \frac{1}{2} \sum_a (z_{ia} - \mu_i - \sum_q f_{aq}\lambda_{iq}-P_{ia})^2\right).
     \end{equation*}
    
    \item Sample \( \bm{\theta}_i = (\mu_i, \bm{\lambda}_i) \) from \( p(\bm{\theta}_i \mid \bm{z}_i, \bm{F}, \sigma_i^2) \) using the proposed PX-DA algorithm:\\[1mm]

    \begin{enumerate} \itemsep 2mm

    \item sample $\delta^* \sim \mathcal{IG}(d_0, D_0)$;
    \item set $\Tilde{z}_{ia} = z_{ia} \sqrt{\delta^*}$ for all 
    {$a \in \{1, \ldots, A\}$} where $y_{ia}=0$, otherwise $\Tilde{z}_{ia} = z_{ia}$;
    \item sample $\delta \sim p(\delta | \bm{\Tilde{z}}_i, \sigma_i^2, \bm{F}, \bm{y}_i)$ using the algorithm given in Appendix~\ref{appA};
    \item set $z_{ia}^{\text{new}} = \Tilde{z}_{ia} / \sqrt{\delta}$ for all {$a \in \{1, \ldots, A\}$} where $y_{ia}=0$, otherwise ${z}^{\text{new}}_{ia} = z_{ia}$.        \item Finally, draw 
            \begin{equation*}
            \begin{aligned}
            \bm{\theta}_i {|\cdot}  &\sim \mathcal{N}(\bm{T}_{N,i}\bm{t}_{N,i}, \bm{T}_{N,i}),\\
            \bm{T}_{N,i}^{-1} &= \bm{T}_0^{-1} + \frac{1}{\sigma_i^2} \bm{W}^\top\bm{W}, \quad \bm{t}_{N,i} = \frac{1}{\sigma_i^2} \bm{W}^\top (\bm{z}^{\text{new}}_i - \bm{P}_i).
            \end{aligned}
            \end{equation*}
    \end{enumerate}
    \end{enumerate}
\end{enumerate}
where $\bm{W} = (\iota_A, \bm{F})$ is the $A \times (Q+1)$ regressor matrix for updating $\bm{\theta}_i$ with rows $\bm{w}_a$,  $\iota_A$ is an $A \times 1$ vector of ones, and $\bm{T}_0 = \text{diag}(M_0, L_0, \dots, L_0)$ is a diagonal matrix of dimension $Q+1$. 

The relevant posterior parameters to sample $\delta \sim p(\delta | \bm{\Tilde{z}}_i, \sigma_i^2, \bm{F}, \bm{y}_i)$ 
{in Step (2c-iii)} are given by 
\begin{equation}
\label{eq:delta_moments_factor}
    \begin{aligned}
      d_I &= d_0 + \frac{n_0}{2},\\
      D_I &= D_0 + \frac{\delta^*}{2\sigma_i^2} \sum_{a:y_{ia}=0} {z_{ia}^2}
      - \frac{\delta^*}{2} \mathbf{m}_0^\top {\mathbf{T}_{N,i}} \mathbf{m}_0,\\
      B_I &= \sqrt{\delta^*} \left(\frac{1}{\sigma^2_i}\sum_{a:y_{ia}=0} (z_{ia} P_{ia}) +  \mathbf{m}_0^\top {\mathbf{T}_{N,i}} (\mathbf{m}_1 - \mathbf{m}_2) \right),
    \end{aligned}
\end{equation}
where $n_0 = \sum_a\mathbbm{1}(y_{ia}=0)$, $\mathbf{m}_0 = \sigma_i^{-2}\sum_{a:y_{ia}=0} \mathbf{w}_a^\top z_{ia}$, $\mathbf{m}_1 = \sigma_i^{-2}\sum_{a:y_{ia}>0} \mathbf{w}_a^\top z_{ia}$, and $\mathbf{m}_2 = \sigma_i^{-2}\sum_a \mathbf{w}_a^\top {P_{ia}}$.
For notational simplicity, we have suppressed dependence on $i$ in several quantities in (\ref{eq:delta_moments_factor}).

\end{appendices}


\begin{thebibliography}{32}
\ifx \bisbn   \undefined \def \bisbn  #1{ISBN #1}\fi
\ifx \binits  \undefined \def \binits#1{#1}\fi
\ifx \bauthor  \undefined \def \bauthor#1{#1}\fi
\ifx \batitle  \undefined \def \batitle#1{#1}\fi
\ifx \bjtitle  \undefined \def \bjtitle#1{#1}\fi
\ifx \bvolume  \undefined \def \bvolume#1{\textbf{#1}}\fi
\ifx \byear  \undefined \def \byear#1{#1}\fi
\ifx \bissue  \undefined \def \bissue#1{#1}\fi
\ifx \bfpage  \undefined \def \bfpage#1{#1}\fi
\ifx \blpage  \undefined \def \blpage #1{#1}\fi
\ifx \burl  \undefined \def \burl#1{\textsf{#1}}\fi
\ifx \doiurl  \undefined \def \doiurl#1{\url{https://doi.org/#1}}\fi
\ifx \betal  \undefined \def \betal{\textit{et al.}}\fi
\ifx \binstitute  \undefined \def \binstitute#1{#1}\fi
\ifx \binstitutionaled  \undefined \def \binstitutionaled#1{#1}\fi
\ifx \bctitle  \undefined \def \bctitle#1{#1}\fi
\ifx \beditor  \undefined \def \beditor#1{#1}\fi
\ifx \bpublisher  \undefined \def \bpublisher#1{#1}\fi
\ifx \bbtitle  \undefined \def \bbtitle#1{#1}\fi
\ifx \bedition  \undefined \def \bedition#1{#1}\fi
\ifx \bseriesno  \undefined \def \bseriesno#1{#1}\fi
\ifx \blocation  \undefined \def \blocation#1{#1}\fi
\ifx \bsertitle  \undefined \def \bsertitle#1{#1}\fi
\ifx \bsnm \undefined \def \bsnm#1{#1}\fi
\ifx \bsuffix \undefined \def \bsuffix#1{#1}\fi
\ifx \bparticle \undefined \def \bparticle#1{#1}\fi
\ifx \barticle \undefined \def \barticle#1{#1}\fi
\bibcommenthead
\ifx \bconfdate \undefined \def \bconfdate #1{#1}\fi
\ifx \botherref \undefined \def \botherref #1{#1}\fi
\ifx \url \undefined \def \url#1{\textsf{#1}}\fi
\ifx \bchapter \undefined \def \bchapter#1{#1}\fi
\ifx \bbook \undefined \def \bbook#1{#1}\fi
\ifx \bcomment \undefined \def \bcomment#1{#1}\fi
\ifx \oauthor \undefined \def \oauthor#1{#1}\fi
\ifx \citeauthoryear \undefined \def \citeauthoryear#1{#1}\fi
\ifx \endbibitem  \undefined \def \endbibitem {}\fi
\ifx \bconflocation  \undefined \def \bconflocation#1{#1}\fi
\ifx \arxivurl  \undefined \def \arxivurl#1{\textsf{#1}}\fi
\csname PreBibitemsHook\endcsname

\bibitem[\protect\citeauthoryear{Gschl{\"o\ss}l and Czado}{2008}]{gsc-cza:mod}
\begin{barticle}
\bauthor{\bsnm{Gschl{\"o\ss}l}, \binits{S.}},
\bauthor{\bsnm{Czado}, \binits{C.}}:
\batitle{Modelling count data with overdispersion and spatial effects}.
\bjtitle{Statistical Papers}
\bvolume{49},
\bfpage{531}--\blpage{552}
(\byear{2008})
\end{barticle}
\endbibitem

\bibitem[\protect\citeauthoryear{Czado et~al.}{2009}]{czado2009predictive}
\begin{barticle}
\bauthor{\bsnm{Czado}, \binits{C.}},
\bauthor{\bsnm{Gneiting}, \binits{T.}},
\bauthor{\bsnm{Held}, \binits{L.}}:
\batitle{{Predictive model assessment for count data}}.
\bjtitle{Biometrics}
\bvolume{65}(\bissue{4}),
\bfpage{1254}--\blpage{1261}
(\byear{2009})
\end{barticle}
\endbibitem

\bibitem[\protect\citeauthoryear{Czado et~al.}{2005}]{czado2005bayesian}
\begin{barticle}
\bauthor{\bsnm{Czado}, \binits{C.}},
\bauthor{\bsnm{Delwarde}, \binits{A.}},
\bauthor{\bsnm{Denuit}, \binits{M.}}:
\batitle{{Bayesian Poisson log-bilinear mortality projections}}.
\bjtitle{Insurance: Mathematics and Economics}
\bvolume{36}(\bissue{3}),
\bfpage{260}--\blpage{284}
(\byear{2005})
\end{barticle}
\endbibitem

\bibitem[\protect\citeauthoryear{Silva and Tenreyro}{2010}]{silva2010existence}
\begin{barticle}
\bauthor{\bsnm{Silva}, \binits{J.S.}},
\bauthor{\bsnm{Tenreyro}, \binits{S.}}:
\batitle{{On the existence of the maximum likelihood estimates in Poisson regression}}.
\bjtitle{Economics Letters}
\bvolume{107}(\bissue{2}),
\bfpage{310}--\blpage{312}
(\byear{2010})
\end{barticle}
\endbibitem

\bibitem[\protect\citeauthoryear{Gamerman}{1997}]{gamerman1997sampling}
\begin{barticle}
\bauthor{\bsnm{Gamerman}, \binits{D.}}:
\batitle{Sampling from the posterior distribution in generalized linear mixed models}.
\bjtitle{Statistics and Computing}
\bvolume{7},
\bfpage{57}--\blpage{68}
(\byear{1997})
\end{barticle}
\endbibitem

\bibitem[\protect\citeauthoryear{Chib and Winkelmann}{2001}]{chib2001markov}
\begin{barticle}
\bauthor{\bsnm{Chib}, \binits{S.}},
\bauthor{\bsnm{Winkelmann}, \binits{R.}}:
\batitle{{Markov chain Monte Carlo analysis of correlated count data}}.
\bjtitle{Journal of Business \& Economic Statistics}
\bvolume{19}(\bissue{4}),
\bfpage{428}--\blpage{435}
(\byear{2001})
\end{barticle}
\endbibitem

\bibitem[\protect\citeauthoryear{Polson et~al.}{2013}]{pol-etal:bay_inf}
\begin{barticle}
\bauthor{\bsnm{Polson}, \binits{N.G.}},
\bauthor{\bsnm{Scott}, \binits{J.G.}},
\bauthor{\bsnm{Windle}, \binits{J.}}:
\batitle{Bayesian inference for logistic models using {{P}\'{o}lya-{G}amma} latent variables}.
\bjtitle{Journal of the American Statistical Association}
\bvolume{108},
\bfpage{1339}--\blpage{49}
(\byear{2013})
\end{barticle}
\endbibitem

\bibitem[\protect\citeauthoryear{Tanner and Wong}{1987}]{tanner1987calculation}
\begin{barticle}
\bauthor{\bsnm{Tanner}, \binits{M.A.}},
\bauthor{\bsnm{Wong}, \binits{W.H.}}:
\batitle{The calculation of posterior distributions by data augmentation}.
\bjtitle{Journal of the American Statistical Association}
\bvolume{82}(\bissue{398}),
\bfpage{528}--\blpage{540}
(\byear{1987})
\end{barticle}
\endbibitem

\bibitem[\protect\citeauthoryear{Fr{\"u}hwirth-Schnatter and Wagner}{2006}]{fruhwirth2006auxiliary}
\begin{barticle}
\bauthor{\bsnm{Fr{\"u}hwirth-Schnatter}, \binits{S.}},
\bauthor{\bsnm{Wagner}, \binits{H.}}:
\batitle{Auxiliary mixture sampling for parameter-driven models of time series of counts with applications to state space modelling}.
\bjtitle{Biometrika}
\bvolume{93}(\bissue{4}),
\bfpage{827}--\blpage{841}
(\byear{2006})
\end{barticle}
\endbibitem

\bibitem[\protect\citeauthoryear{Fr{\"u}hwirth-Schnatter et~al.}{2009}]{fruhwirth2009improved}
\begin{barticle}
\bauthor{\bsnm{Fr{\"u}hwirth-Schnatter}, \binits{S.}},
\bauthor{\bsnm{Fr{\"u}hwirth}, \binits{R.}},
\bauthor{\bsnm{Held}, \binits{L.}},
\bauthor{\bsnm{Rue}, \binits{H.}}:
\batitle{{Improved auxiliary mixture sampling for hierarchical models of non-Gaussian data}}.
\bjtitle{Statistics and Computing}
\bvolume{19},
\bfpage{479}--\blpage{492}
(\byear{2009})
\end{barticle}
\endbibitem

\bibitem[\protect\citeauthoryear{Pillow and Scott}{2012}]{pillow2012fully}
\begin{botherref}
\oauthor{\bsnm{Pillow}, \binits{J.}},
\oauthor{\bsnm{Scott}, \binits{J.}}:
Fully {B}ayesian inference for neural models with negative-binomial spiking.
Advances in Neural Information Processing Systems
\textbf{25}
(2012)
\end{botherref}
\endbibitem

\bibitem[\protect\citeauthoryear{Steel and Zens}{Forthcoming}]{steel2024model}
\begin{botherref}
\oauthor{\bsnm{Steel}, \binits{M.F.}},
\oauthor{\bsnm{Zens}, \binits{G.}}:
{Model Uncertainty in Latent Gaussian Models with Univariate Link Function}.
Bayesian Analysis
(Forthcoming)
\end{botherref}
\endbibitem

\bibitem[\protect\citeauthoryear{Canale and Dunson}{2013}]{canale2013nonparametric}
\begin{barticle}
\bauthor{\bsnm{Canale}, \binits{A.}},
\bauthor{\bsnm{Dunson}, \binits{D.B.}}:
\batitle{{Nonparametric Bayes modelling of count processes}}.
\bjtitle{Biometrika}
\bvolume{100}(\bissue{4}),
\bfpage{801}--\blpage{816}
(\byear{2013})
\end{barticle}
\endbibitem

\bibitem[\protect\citeauthoryear{Kowal and Canale}{2020}]{kowal2020simultaneous}
\begin{barticle}
\bauthor{\bsnm{Kowal}, \binits{D.R.}},
\bauthor{\bsnm{Canale}, \binits{A.}}:
\batitle{{Simultaneous transformation and rounding (STAR) models for integer-valued data}}.
\bjtitle{Electronic Journal of Statistics}
\bvolume{14}(\bissue{1}),
\bfpage{1744}--\blpage{1772}
(\byear{2020})
\end{barticle}
\endbibitem

\bibitem[\protect\citeauthoryear{Meng and Van~Dyk}{1999}]{meng1999seeking}
\begin{barticle}
\bauthor{\bsnm{Meng}, \binits{X.-L.}},
\bauthor{\bsnm{Van~Dyk}, \binits{D.A.}}:
\batitle{Seeking efficient data augmentation schemes via conditional and marginal augmentation}.
\bjtitle{Biometrika}
\bvolume{86}(\bissue{2}),
\bfpage{301}--\blpage{320}
(\byear{1999})
\end{barticle}
\endbibitem

\bibitem[\protect\citeauthoryear{Duan et~al.}{2018}]{duan2018scaling}
\begin{barticle}
\bauthor{\bsnm{Duan}, \binits{L.L.}},
\bauthor{\bsnm{Johndrow}, \binits{J.E.}},
\bauthor{\bsnm{Dunson}, \binits{D.B.}}:
\batitle{{Scaling up data augmentation MCMC via calibration}}.
\bjtitle{Journal of Machine Learning Research}
\bvolume{19}(\bissue{64}),
\bfpage{1}--\blpage{34}
(\byear{2018})
\end{barticle}
\endbibitem

\bibitem[\protect\citeauthoryear{Zens et~al.}{2024}]{zens2024ultimate}
\begin{barticle}
\bauthor{\bsnm{Zens}, \binits{G.}},
\bauthor{\bsnm{Fr{\"u}hwirth-Schnatter}, \binits{S.}},
\bauthor{\bsnm{Wagner}, \binits{H.}}:
\batitle{{Ultimate P{\'o}lya Gamma Samplers--Efficient MCMC for possibly imbalanced binary and categorical data}}.
\bjtitle{Journal of the American Statistical Association}
\bvolume{119}(\bissue{548}),
\bfpage{2548}--\blpage{2559}
(\byear{2024})
\end{barticle}
\endbibitem

\bibitem[\protect\citeauthoryear{Liu and Wu}{1999}]{liu1999parameter}
\begin{barticle}
\bauthor{\bsnm{Liu}, \binits{J.S.}},
\bauthor{\bsnm{Wu}, \binits{Y.N.}}:
\batitle{Parameter expansion for data augmentation}.
\bjtitle{Journal of the American Statistical Association}
\bvolume{94}(\bissue{448}),
\bfpage{1264}--\blpage{1274}
(\byear{1999})
\end{barticle}
\endbibitem

\bibitem[\protect\citeauthoryear{Van~Dyk and Meng}{2001}]{van2001art}
\begin{barticle}
\bauthor{\bsnm{Van~Dyk}, \binits{D.A.}},
\bauthor{\bsnm{Meng}, \binits{X.-L.}}:
\batitle{The art of data augmentation}.
\bjtitle{Journal of Computational and Graphical Statistics}
\bvolume{10}(\bissue{1}),
\bfpage{1}--\blpage{50}
(\byear{2001})
\end{barticle}
\endbibitem

\bibitem[\protect\citeauthoryear{Plummer et~al.}{2006}]{codaR}
\begin{barticle}
\bauthor{\bsnm{Plummer}, \binits{M.}},
\bauthor{\bsnm{Best}, \binits{N.}},
\bauthor{\bsnm{Cowles}, \binits{K.}},
\bauthor{\bsnm{Vines}, \binits{K.}}, \betal:
\batitle{{CODA: convergence diagnosis and output analysis for MCMC}}.
\bjtitle{R News}
\bvolume{6}(\bissue{1}),
\bfpage{7}--\blpage{11}
(\byear{2006})
\end{barticle}
\endbibitem

\bibitem[\protect\citeauthoryear{Raftery et~al.}{2012}]{raftery2012bayesian}
\begin{barticle}
\bauthor{\bsnm{Raftery}, \binits{A.E.}},
\bauthor{\bsnm{Li}, \binits{N.}},
\bauthor{\bsnm{{\v{S}}ev{\v{c}}{\'\i}kov{\'a}}, \binits{H.}},
\bauthor{\bsnm{Gerland}, \binits{P.}},
\bauthor{\bsnm{Heilig}, \binits{G.K.}}:
\batitle{Bayesian probabilistic population projections for all countries}.
\bjtitle{Proceedings of the National Academy of Sciences}
\bvolume{109}(\bissue{35}),
\bfpage{13915}--\blpage{13921}
(\byear{2012})
\end{barticle}
\endbibitem

\bibitem[\protect\citeauthoryear{Raftery et~al.}{2014}]{raftery2014bayesian}
\begin{barticle}
\bauthor{\bsnm{Raftery}, \binits{A.E.}},
\bauthor{\bsnm{Alkema}, \binits{L.}},
\bauthor{\bsnm{Gerland}, \binits{P.}}:
\batitle{{Bayesian population projections for the United Nations}}.
\bjtitle{Statistical Science}
\bvolume{29}(\bissue{1}),
\bfpage{58}
(\byear{2014})
\end{barticle}
\endbibitem

\bibitem[\protect\citeauthoryear{Bijak and Bryant}{2016}]{bijak2016bayesian}
\begin{barticle}
\bauthor{\bsnm{Bijak}, \binits{J.}},
\bauthor{\bsnm{Bryant}, \binits{J.}}:
\batitle{{Bayesian demography 250 years after Bayes}}.
\bjtitle{Population Studies}
\bvolume{70}(\bissue{1}),
\bfpage{1}--\blpage{19}
(\byear{2016})
\end{barticle}
\endbibitem

\bibitem[\protect\citeauthoryear{Wi{\'s}niowski et~al.}{2015}]{wisniowski2015bayesian}
\begin{barticle}
\bauthor{\bsnm{Wi{\'s}niowski}, \binits{A.}},
\bauthor{\bsnm{Smith}, \binits{P.W.}},
\bauthor{\bsnm{Bijak}, \binits{J.}},
\bauthor{\bsnm{Raymer}, \binits{J.}},
\bauthor{\bsnm{Forster}, \binits{J.J.}}:
\batitle{{Bayesian population forecasting: extending the Lee-Carter method}}.
\bjtitle{Demography}
\bvolume{52}(\bissue{3}),
\bfpage{1035}--\blpage{1059}
(\byear{2015})
\end{barticle}
\endbibitem

\bibitem[\protect\citeauthoryear{Alexander et~al.}{2017}]{alexander2017flexible}
\begin{barticle}
\bauthor{\bsnm{Alexander}, \binits{M.}},
\bauthor{\bsnm{Zagheni}, \binits{E.}},
\bauthor{\bsnm{Barbieri}, \binits{M.}}:
\batitle{{A flexible Bayesian model for estimating subnational mortality}}.
\bjtitle{Demography}
\bvolume{54}(\bissue{6}),
\bfpage{2025}--\blpage{2041}
(\byear{2017})
\end{barticle}
\endbibitem

\bibitem[\protect\citeauthoryear{Zens}{forthcoming}]{zens2024flexible}
\begin{botherref}
\oauthor{\bsnm{Zens}, \binits{G.}}:
{Flexible Bayesian Modeling of Age-Specific Counts in Many Demographic Subpopulations}.
Journal of the Royal Statistical Society: Series A (Statistics in Society)
(forthcoming)
\end{botherref}
\endbibitem

\bibitem[\protect\citeauthoryear{Lee and Carter}{1992}]{lee1992modeling}
\begin{barticle}
\bauthor{\bsnm{Lee}, \binits{R.D.}},
\bauthor{\bsnm{Carter}, \binits{L.R.}}:
\batitle{{Modeling and forecasting US mortality}}.
\bjtitle{Journal of the American Statistical Association}
\bvolume{87}(\bissue{419}),
\bfpage{659}--\blpage{671}
(\byear{1992})
\end{barticle}
\endbibitem

\bibitem[\protect\citeauthoryear{Clark}{2019}]{clark2019general}
\begin{barticle}
\bauthor{\bsnm{Clark}, \binits{S.J.}}:
\batitle{{A general age-specific mortality model with an example indexed by child mortality or both child and adult mortality}}.
\bjtitle{Demography}
\bvolume{56}(\bissue{3}),
\bfpage{1131}--\blpage{1159}
(\byear{2019})
\end{barticle}
\endbibitem

\bibitem[\protect\citeauthoryear{Vehtari et~al.}{2017}]{vehtari2017practical}
\begin{barticle}
\bauthor{\bsnm{Vehtari}, \binits{A.}},
\bauthor{\bsnm{Gelman}, \binits{A.}},
\bauthor{\bsnm{Gabry}, \binits{J.}}:
\batitle{{Practical Bayesian model evaluation using leave-one-out cross-validation and WAIC}}.
\bjtitle{Statistics and Computing}
\bvolume{27},
\bfpage{1413}--\blpage{1432}
(\byear{2017})
\end{barticle}
\endbibitem

\bibitem[\protect\citeauthoryear{Chen and Li}{2022}]{chen2022determining}
\begin{barticle}
\bauthor{\bsnm{Chen}, \binits{Y.}},
\bauthor{\bsnm{Li}, \binits{X.}}:
\batitle{Determining the number of factors in high-dimensional generalized latent factor models}.
\bjtitle{Biometrika}
\bvolume{109}(\bissue{3}),
\bfpage{769}--\blpage{782}
(\byear{2022})
\end{barticle}
\endbibitem

\bibitem[\protect\citeauthoryear{Mauri and Dunson}{2024}]{mauri2024factor}
\begin{botherref}
\oauthor{\bsnm{Mauri}, \binits{L.}},
\oauthor{\bsnm{Dunson}, \binits{D.B.}}:
{Factor pre-training in Bayesian multivariate logistic models}.
arXiv preprint arXiv:2409.17441
(2024)
\end{botherref}
\endbibitem

\bibitem[\protect\citeauthoryear{Liu et~al.}{1998}]{liu1998parameter}
\begin{botherref}
\oauthor{\bsnm{Liu}, \binits{C.}},
\oauthor{\bsnm{Rubin}, \binits{D.B.}},
\oauthor{\bsnm{Wu}, \binits{Y.N.}}:
{Parameter expansion to accelerate EM: the PX-EM algorithm}.
Biometrika,
755--770
(1998)
\end{botherref}
\endbibitem

\end{thebibliography}
\end{document}